\documentclass[aps,prl,preprint,superscriptaddress,showpacs]{revtex4-1}

\usepackage{graphicx}
\usepackage{dcolumn}
\usepackage{bm}

\begin{document}


\title{Guided propagation of extremely intense lasers in plasma via ion motion}

\author{Wei-Min Wang}
\affiliation{Department of Physics
and Beijing Key Laboratory of Opto-electronic Functional Materials
and Micro-nano Devices, Renmin University of China, Beijing 100872,
China}

\author{Zheng-Ming Sheng}
\affiliation{SUPA, Department of Physics, University of Strathclyde,
Glasgow G4 0NG, United Kingdom} \affiliation{Key Laboratory for
Laser Plasmas (MoE) and School of Physics and Astronomy, Shanghai
Jiao Tong University, Shanghai 200240, China}\affiliation{Tsung-Dao
Lee Institute, Shanghai Jiao Tong University, Shanghai 200240,
China}

\author{Thomas Wilson}
\affiliation{SUPA, Department of Physics, University of Strathclyde,
Glasgow G4 0NG, United Kingdom}

\author{Yu-Tong Li}
\affiliation{Beijing National Laboratory for Condensed Matter
Physics, Institute of Physics, CAS, Beijing 100190, China}
\affiliation{School of Physical Sciences, University of Chinese
Academy of Sciences, Beijing 100049, China} \affiliation{Songshan
Lake Materials Laboratory, Dongguan, Guangdong 523808, China}

\author{Jie Zhang}
\affiliation{Beijing National Laboratory for Condensed Matter
Physics, Institute of Physics, CAS, Beijing 100190, China}
\affiliation{Key Laboratory for Laser Plasmas (MoE) and School of
Physics and Astronomy, Shanghai Jiao Tong University, Shanghai
200240, China}

\date{\today}

\begin{abstract}
The upcoming $10-100$ petawatt laser facilities may deliver laser
pulses with unprecedented intensity of $10^{22}-10^{25}\rm~W
cm^{-2}$, which can trigger various nonlinear quantum electrodynamic
processes in plasma. For effective laser plasma interactions at such
high intensity levels, guided laser propagation is critical.
However, this becomes impossible via usual plasma electron response
to laser fields due to electron cavitation by the laser
ponderomotive force. Here, we find that ion response to the laser
fields may effectively guide laser propagation at such high
intensity levels. The corresponding conditions of the required ion
density distribution and laser power are presented and verified by
three-dimensional particle-in-cell simulations. Our theory shall
serve as a guide for future experimental design involving ultrahigh
intensity lasers.
\end{abstract}


\maketitle

With the new progresses in high power laser technologies, laser
pulses with the peak powers of multi-petawatts (PW) are becoming
available recently
\cite{Danson,Laser_4PW20fs,Laser_1PW_contrast,SULF_5PW}. Even higher
power laser systems at the 100 PW level are planned or under
construction \cite{ELI,OMEGA_EP_OPAL,XCELS}. With these systems one
may achieve focused laser intensity at the unprecedented level of
$10^{22}-10^{25}\rm~W cm^{-2}$. This enables one to explore
fundamental physics and applications at high intensity frontiers
such as quantum electrodynamics (QED) effects in plasma and vacuum
\cite{Bell,Piazza,Ridgers,Vacuum_QED}, electron acceleration over 10
GeV \cite{Tajima,Esarey,Lu,Leemans_2014,Leemans_2019}, GeV ion
acceleration \cite{Esirkepov,Yan,Bulanov,Macchi,Mima,Matsui}, and
high-energy gamma-ray generation \cite{PNAS2018,Lei,Chen}, etc.

To achieve and maintain extremely high laser intensity within an
enough distance along the laser propagation direction for the above
applications, guided laser propagation is critical. Guided laser
propagation in a plasma channel \cite{Esarey97,Channel_exp} or in
uniform plasma \cite{Monot,Pukhov} at the laser intensity less than
$10^{20} {\rm W/cm^2}$ has been well demonstrated and adopted in
applications such as laser wakefield acceleration
\cite{Leemans_2014,Leemans_2019}. This is achieved via linear and
nonlinear plasma electron response to the laser fields. At higher
intensity, however, the strong laser ponderomotive force can
significantly expel electrons from the laser interaction zone
\cite{Sun87,Kneip,Wang2012} and therefore, the normal channel
guiding \cite{Esarey97,Channel_exp} and relativistic self-guiding
\cite{Monot,Pukhov} via the electron response do not work anymore
\cite{Wang2012}. Hence, it becomes challenging to achieve guided
laser propagation with laser intensity above $ 10^{20}\rm~W
cm^{-2}$.

In this paper, we show that both the channel guiding and
relativistic self-guiding can be achieved \emph{via ion response} to
the laser fields even if complete electron cavitation occurs along
the laser axis. When the ion density has a proper transverse density
profile to provide the refractive index peaked along the channel
axis, channel guiding can still occur via linear ion response. When
the laser power exceeds a certain threshold, relativistic
self-guiding due to nonlinear ion response also develops. The
criterions for the two kinds of guiding are identified and then
verified by three-dimensional (3D) particle-in-cell (PIC)
simulations.

\begin{figure}[htbp]
\includegraphics[width=3.5in]{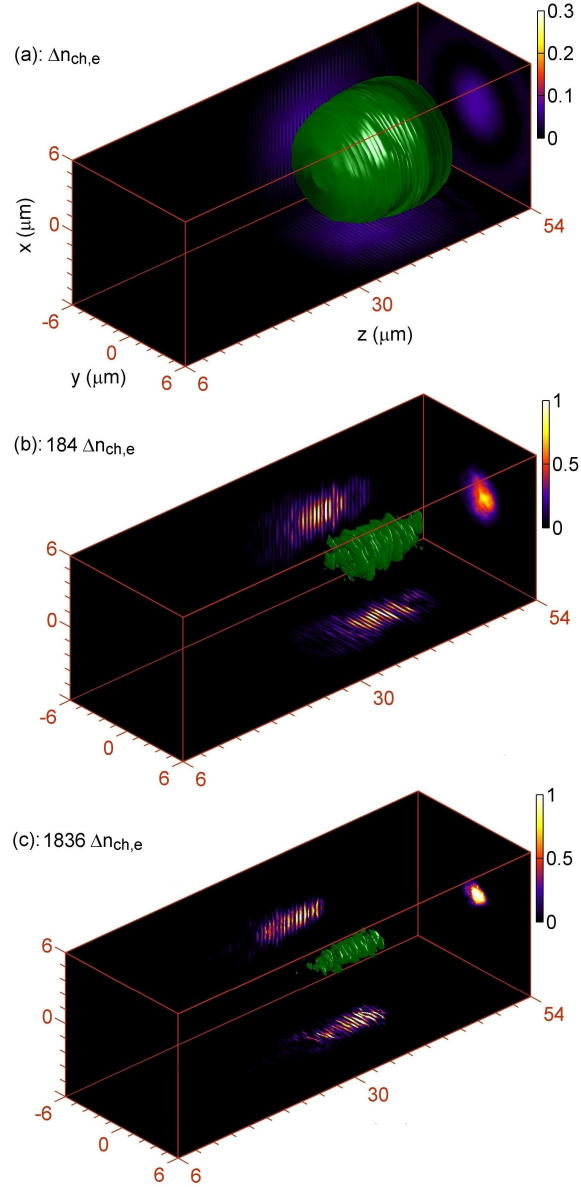}
\caption{\label{fig:epsart} Three-dimensional isosurfaces of the
laser intensity $I/I_0$ ($I_0$ is the initial intensity) as well as
the slices at the planes with respective peak values at the time 50
$\tau_0$ or 4 $z_R/c$. The plots  (a), (b) and (c) correspond to
different channel depths of $\triangle n_{ch,e}$, $184\triangle
n_{ch,e}$, and $1836\triangle n_{ch,e}$, respectively, where
$\triangle n_{ch,e}$ is the normal critical density depth due to the
electron response.}
\end{figure}

We first present examples to show how to obtain guided propagation
of an ultrahigh intensity laser pulse in plasma through 3D PIC
simulations with the KLAPS code \cite{KLAPS, Wang-PRE17}. Usually,
to achieve enough high intensities up to $10^{22}-10^{25}\rm~W
cm^{-2}$, tightly focusing with a spot radius down to few
wavelengths is essential. In our simulations, we adopt the spot
radius $r_0=2.0\rm\mu m$. A laser pulse propagates along the $+z$
direction with the $x$-direction polarization, a wavelength
$\lambda_0=1\rm\mu m$ (laser period $\tau_0=2\pi/\omega_0=3.33\rm
fs$), amplitude $a_0=200$ normalized by $m_ec\omega_0/e$ (the
corresponding intensity $5.5\times 10^{22}\rm~W cm^{-2}$), and
duration 40fs in full width at half maximum (FWHM). Here, $e$ and
$m_e$ are the electron charge and mass, and $c$ is the light speed
in vacuum. The laser pulse peak arrives at the left boundary of a
plasma at 12 $\tau_0$. A preformed plasma channel is taken with a
parabolic density profile $n=n_0+\triangle n\times r^2/r_0^2$ within
$r\leq 2r_0$ and $n=n_0+4\triangle n$ within $r> 2r_0$, where
$\triangle n$ is the channel depth. The plasma channel is composed
of electrons and protons. We adopt a moving window at the light
speed $c$. The window has a simulation box $12 \mu m \times 12 \mu m
\times 96 \mu m$ in $x\times y \times z$ directions (or $24 \mu m
\times 24 \mu m \times 48 \mu m$ in the case when laser defocusing
occurs). The resolutions along the $z$ and two transverse directions
are 0.02$\rm\mu m$ and 0.25$\rm\mu m$, respectively. Eight
simulation electrons and ions are taken per cell.

Under different plasma channel parameters, Fig. 1 shows laser
intensity profiles after the propagation of 4 $z_R$ ($z_R=\pi
r_0^2/\lambda_0$ is the Rayleigh length). Figure 1(a) shows that the
laser pulse of $a_0=200$ cannot be guided by a plasma channel with a
density depth $\triangle n=\triangle n_{ch,e}$, where $\triangle
n_{ch,e}=m_ec^2/(\pi r_0^2 e^2)\simeq 0.1 n_{c,e}\times(1\mu
m/r_0)^2$ is the critical depth determined by the electron response
\cite{Esarey97} and $n_{c,e}=m_e \omega_0^2/4\pi
e^2=1.1\times10^{21} \rm cm^{-3} \times (1\mu m/\lambda_0)^2$ is the
critical density of plasma electrons. Experiments and our
simulations have showed that such a channel can well guide a laser
pulse of $a_0 \sim 1$ over many Rayleigh lengths
\cite{Channel_exp,Leemans_2014} because a transverse profile of the
refractive index peaked along the channel axis \cite{Esarey97}.
However, with the high amplitude of $a_0=200$, the laser pulse can
quickly push the plasma electrons away from its interaction zone by
its ponderomotive force, as illustrated in Fig. 2(a). Therefore, the
refractive index profile suitable for laser guiding disappears.

\begin{figure}[htbp]
\includegraphics[width=4.5in]{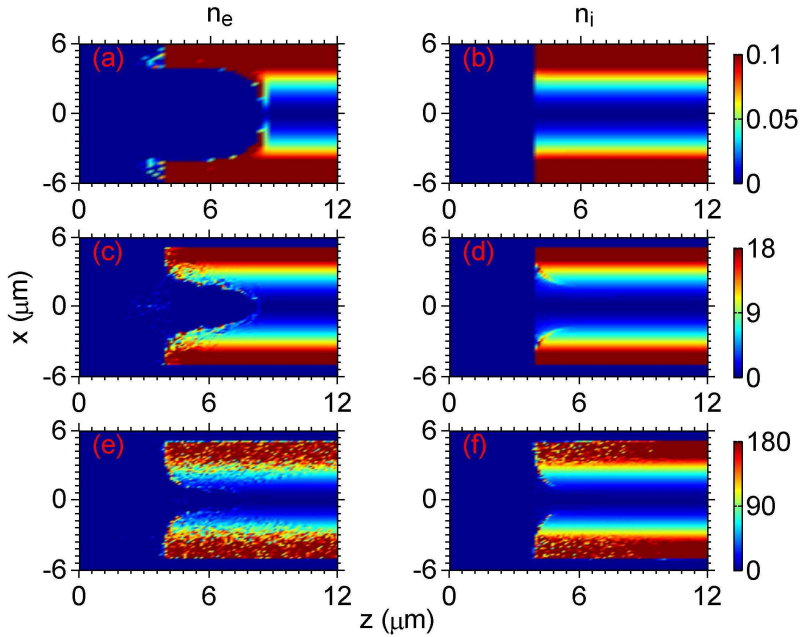}
\caption{\label{fig:epsart} Electron (left) and ion (right) density
distributions at 10$\tau_0$, where the channel depth is taken as
$\triangle n_{ch,e}$ in (a,b), $184\triangle n_{ch,e}$ in (c, d),
and $1836\triangle n_{ch,e}$ in (e, f). }
\end{figure}

As the density depth is increased to $\Delta n=184\triangle
n_{ch,e}$ and $1836\triangle n_{ch,e}$ in Figs. 1(b) and 1(c),
respectively, the laser pulses are guided better. In particular,
with $\Delta n=1836\triangle n_{ch,e}$, the laser spot radius is
kept around its initial value $r_0=2.0\rm\mu m$ over a few Rayleigh
lengths [see Figs. 1(c)]. One could explain this result as ions with
higher densities tend to prevent the expulsion of electrons from the
channel axis and the remaining electrons can help the laser
focusing. On the other hand, we find that ion motion is key in this
case.

\begin{figure}[htbp]
\includegraphics[width=4in]{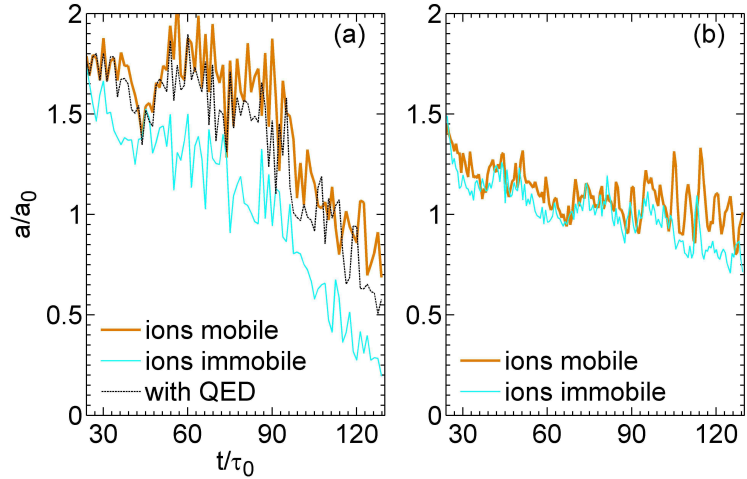}
\caption{\label{fig:epsart}Evolution of the laser amplitude peak
when the channel depth is taken as $1836\triangle n_{ch,e}$ in (a)
and $184\triangle n_{ch,e}$ in (b). Different curves correspond to
simulations with QED effects, and ions mobile or immobile. }
\end{figure}

We set the ions immobile in our simulation shown in the cyan line in
Fig. 3(a). The evolution of the laser amplitude shows obvious
difference from the case with the ions mobile (dark-orange line).
Furthermore, immobile ions result in weaker laser focusing.
Normally, immobile ions tend to prevent the electron expulsion and
thus retain the electron density profile better. Hence, immobile
ions should have caused stronger laser focusing. However, these
results can be explained by ion response to the laser fields.
Actually, Fig. 2 show that while electrons are quickly expelled from
the channel axis, the high mass of the ions slows their response.
Ions remain near their initial locations much longer, eventually
following the electrons in a delayed fashion. The oscillation of
these ions in the laser fields can lead to the refractive index
distribution peaked at the channel axis, which can help laser
focusing.

According to the dispersion relation of a planar laser pulse in
plasma, i.e.,
$\omega_0^2=k^2c^2+\omega_{p,e}^2/\gamma_e+\omega_{p,i}^2/\gamma_i$,
the refractive index can be derived as
\begin{eqnarray}
\eta \simeq  1-\frac{n_e}{2\gamma_e n_{c,e}}-\frac{n_i}{2\gamma_i
n_{c,e}}\frac{m_e q^2}{m_i e^2},
\end{eqnarray}
where $\omega_{p,e}=\sqrt{4\pi n_{e}e^2/m_e}$ and
$\omega_{p,i}=\sqrt{4\pi n_{i}q^2/m_i}$ are the plasma electron and
ion frequencies, $n_e$ and $n_i$ are the electron and ion densities,
$\gamma_e$ and $\gamma_i$ are the electron and ion Lorentz factors,
and $q$ and $m_i$ are the ion charge and mass. To obtain Eq. (1), we
have taken $n_e\ll \gamma_e n_{c,e} \sim a_0 n_{c,e}$ and $n_i\ll
(m_ie^2/m_eq^2) n_{c,e}$. From Eq. (1), one can obtain
\begin{eqnarray}
\frac{\partial \eta}{\partial r} \simeq - \frac{1}{2 n_{c,e}} \left[
\frac{\partial (n_e/\gamma_e)}{\partial r}+\frac{m_e q^2}{m_i
e^2}\frac{\partial (n_i/\gamma_i)}{\partial r} \right],
\end{eqnarray}
where the first and second terms (defined as $\partial \eta/\partial
r|_e$ and $\partial \eta/\partial r|_i$) on the right-hand side
comes from the electron and ion response, respectively. With a
plasma channel as used by us, both $\partial \eta/\partial r|_e$ and
$\partial \eta/\partial r|_i$ are negative initially and $\eta$ has
a peak at the channel axis. Usually the ion contribution can be
ignored because $|\partial \eta/\partial r|_i$ is at a level of
$(m_e/m_i)\times|\partial \eta/\partial r|_e$. However, as the
expulsion of electrons becomes stronger, $|\partial \eta/\partial
r|_i$ can gradually exceed $|\partial \eta/\partial r|_e$. In
particular, when full electron cavitation occurs, $|\partial
\eta/\partial r|_e$ is nearly vanished around the laser interaction
zone and only the ion response works. To achieve laser focusing,
$|\partial \eta/\partial r|_i$ should enhance by a level of
$m_i/m_e$ (1836 for protons), which can be realized by increase the
channel density depth $\triangle n$. This can explain Fig. 1(c) that
as $\Delta n$ is increased from $\triangle n_{ch,e}$ to
$1836\triangle n_{ch,e}$, the channel guiding appears again. Note
that with an insufficient high $\Delta n$ of $184\triangle
n_{ch,e}$, the ion response effect is not enough to cause full
channel guiding [Fig. 1(b)]. Also, this effect does not lead to a
significant difference between the cases of mobile and immobile ions
[Fig. 3(b)].

One can notice that the laser amplitude appears a faster decay after
90 $\tau_0$ in Fig. 3(a) than in Fig. 3(b), which is due to stronger
depletion of the laser energy in a higher plasma density. We also
check two QED effects (the nonlinear Compton scattering and
Breit-Wheeler process for pair creation)
\cite{Bell,Piazza,Ridgers,Wang-PRE17}. The black line in Fig. 3(a)
suggests these effects can be ignored when $a_0\leq 200$, in
agreement with Ref. \cite{Zhidkov}. It is worthwhile to point out
that the channel-guiding effect with ion response can be found even
when $a_0$ is decreased to 50. With further decreased $a_0$, the
electron expulsion becomes weak and the laser pulse interacts mostly
with the electrons.

To quantitatively obtain conditions of the channel and self-guiding
due to the ion response, we present a theory in a pure ion
environment with a life-period longer than the laser pulse duration.
As mentioned above, such environment with full electron cavitation
around the laser axis can be formed in the laser interaction under
certain laser intensities \cite{Sun87,Wang2012,Kneip}. Also, it may
be formed by a relativistic high-current electron beam in plasma
\cite{Whittum,Werner}. In this scenario, the laser pulse propagation
is mainly governed by ion motion. One can use the equation for the
laser envelope under the paraxial approximation,
\begin{eqnarray}
\left(\nabla^{2}_{\bot}-\frac{2i\omega_{0}}{c}\frac{\partial}{\partial
z}\right)A_{s,i}=\frac{\omega^{2}_{p,i0}}{c^2}\left(\frac{n_{i}}{\gamma_i
n_{0}}-1 \right)A_{s,i},
\end{eqnarray}
which is similar to that with electron motion considered
\cite{Sun87,ChenXL,Esarey94,Esarey97,Wang2006}. Here, $A_{s,i}$
normalized by $m_ic^2/e$ is the slowly varying envelope of the laser
vector potential, i.e., $A=A_{s,i}\exp [i\omega_{0}(t-z/c)]$, the
ion channel is taken as a parabolic profile $n_{i0}=n_{0}+\triangle
n r^2/r_0^2$,
and $\gamma_i=\sqrt{1+|A_{s,i}|^2/2}$ for a linearly polarized laser
pulse. We consider the rarefied density with
$\omega_{0}^2\gg\omega_{p,i}^2$ and weakly relativistic ion motion
with $|A_{s,i}|^2\ll 1$ and $\gamma_i\approx 1+|A_{s,i}|^2/4$. With
$|A_{s,i}|^2\ll 1$, the ion density perturbation directly by the
laser ponderomotive force can be ignored, i.e., $n_i \simeq n_{i0}$.
To derive an evolution equation for the laser spot radius $r_s$, one
can take the source-dependent expansion method
\cite{Sprangle87,Esarey94,Esarey97} and assume that the laser field
could be adequately approximated by the lowest order Gaussian mode
$A_{s,i}=A_{s,i0}(r_0/r_s)\exp(-r^2/r_s^2)$. One can derive the
evolution of the normalized spot size $R=r_s/r_0$ satisfying
\begin{equation}
\frac{d ^2 R}{d z^2}=\frac{1}{R^3 z_R^2} \left( 1-\frac{P}{P_{c,i}}-
R^4 \frac{\triangle n}{\triangle n_{ch,i}}\right ),
\end{equation}
where the second term on the right-hand side comes from nonlinear
ion motion and the last term is due to ion-density channel. The
critical power for relativistic self-focusing with a uniform density
(taking $\triangle n=0$) is given by
\begin{eqnarray}
P_{c,i}=\frac{m_i^3\omega_0^2c^5}{2\pi n_0 q^4}=(\frac{m_i^3
e^4}{m_e^3 q^4})\times \frac{n_{c,e}}{n_{0}} \times 17.4 ~\rm GW.
\end{eqnarray}
Usually $P_{c,i}$ is a large value, e.g., for protons with $n_0=100
n_{c,e}=0.055n_{c,i}$, $P_{c,i}=1077$ PW well above the current
laser technical capability, where $n_{c,i}=(m_ie^2/m_eq^2)\times
n_{c,e}$ is the critical density of ions. Therefore, channeling
guiding is a more feasible than self-guiding for 10 to 100 PW laser
pulses available currently and in the near future.

The critical channel depth can be given by
\begin{eqnarray}
\triangle n_{ch,i}=\frac{m_ic^2}{\pi r_0^2 q^2}=
\frac{m_ie^2}{m_eq^2} \times \triangle n_{ch,e}.
\end{eqnarray}
For protons $\triangle n_{ch,i}= 1836 \triangle n_{ch,e}\simeq 47
n_{c,e}$ with $r_0=2 \rm \mu m$. This is the reason why the laser
pulse can be better guided by the channel with $\triangle n= 1836
\triangle n_{ch,e}=\triangle n_{ch,i}$ than the ones with $\triangle
n= 184 \triangle n_{ch,e}$ and $\triangle n_{ch,e}$, as shown in
Fig. 1.

\begin{figure}[htbp]
\includegraphics[width=3.5in]{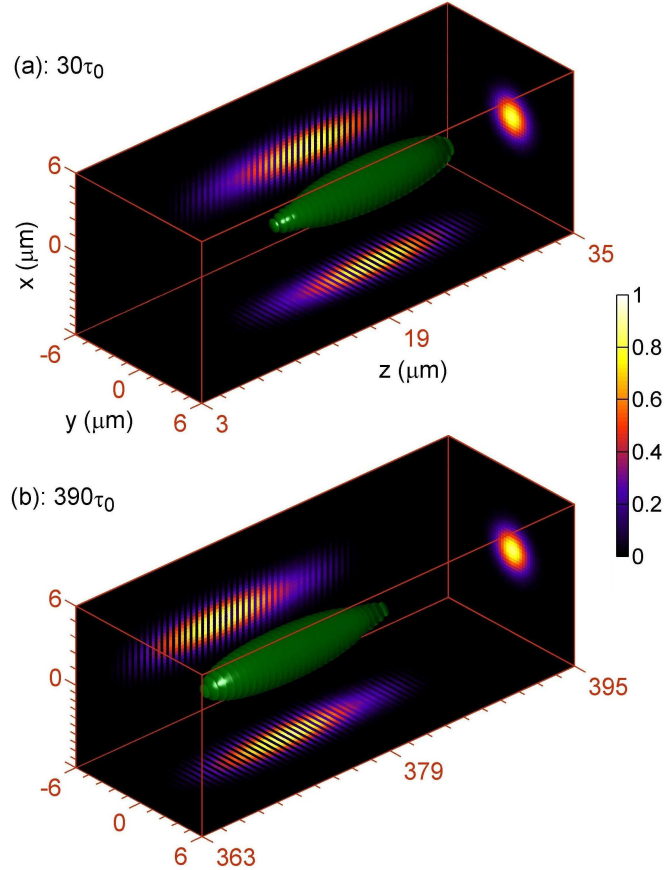}
\caption{\label{fig:epsart}Three-dimensional isosurfaces of the
laser intensity $I/I_0$ ($I_0$ is the initial intensity) as well as
the slices at the planes with respective peak values at the times of
30 $\tau_0$ and 390 $\tau_0$ (31$z_R/c$) in (a) and (b),
respectively.}
\end{figure}

\begin{figure}[htbp]
\includegraphics[width=4in]{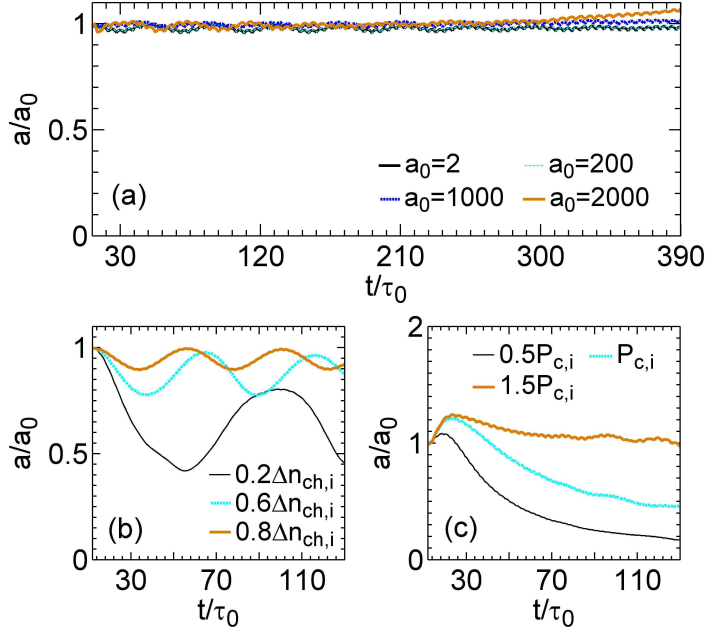}
\caption{\label{fig:epsart}Evolution of the laser amplitude peak.
(a) Different initial amplitudes $a_0$ are taken with a fixed
$\triangle n=\triangle n_{ch,i}$. (b) Different channel depths
$\Delta n$ are taken with a fixed $a_0=200$. (c) Different laser
powers are taken with an initial spot radius $r_0=2\rm \mu m$ and a
uniform density of $100 n_{c,e}=0.055n_{c,i}$ ($n_{c,i}$ is the
critical density of ions).}
\end{figure}

We carry out 3D PIC simulations to test the predicted critical
channel depth and laser power. In Fig. 4, we adopt a proton channel
with $\triangle n=\triangle n_{ch,i}$. It is shown that the laser
pulse is well guided by the channel over 30 Rayleigh lengths. The
channel guiding works in a large range of laser amplitude from
$a_0=2$ to $a_0=2000$ (normalized by $m_ec\omega_0/e$), as shown in
Fig. 5(a). The evolution of the laser amplitudes is almost the same
with $a_0=2$ and $a_0=200$ (the linear ion response is dominant) and
it appears little difference when $a_0$ is increased to 2000, in
which the ion oscillation velocity is close to $c$ and relativistic
effects work. In Fig. 5(b) we decrease $\Delta n$ to $0.8\triangle
n_{ch,i}$, $0.6\triangle n_{ch,i}$, and $0.2\triangle n_{ch,i}$,
respectively, and the channel guiding becomes weaker and even
disappears. These results are in good agreement with Eq. (6).

To examine the critical power given in Eq. (5), we take a uniform
proton density of $100 n_{c,e}=0.055n_{c,i}$. Figure 5(c) shows that
the derived $P_{c,i}$ is valid. When the power is less than
$P_{c,i}$, the laser amplitude decays quickly because the laser
pulse spreads out transversely. When the power is higher than
$P_{c,i}$, self-focusing indeed occurs.

In summary, we find that a ultrahigh intensity laser pulse cannot be
guided either by an underdense plasma channel or by relativistic
nonlinearity due to the electron cavitation formed around the laser
propagation axis. In this case, ion motion becomes important even
with the intensity around $ 10^{22}\rm~W cm^{-2}$. Ion response to
the laser fields can cause effective guiding of such a laser pulse
in certain conditions. A new critical channel depth $\triangle
n_{ch,i}$ and a new critical power $P_{c,i}$ are derived for channel
guiding and self-guiding, respectively, based upon the ion response.
Our 3D PIC simulations show that $\triangle n_{ch,i}$ as the
channel-guiding threshold starts to work when the laser intensity is
sufficient high (e.g., $a_0>50$) and significant cavitation of
electrons occurs. With complete cavitation (free of electrons in the
laser interaction zone), both $\triangle n_{ch,i}$ and $P_{c,i}$ as
guiding thresholds are very accurate. In particular, an ion channel
can stably guide laser pulses with amplitudes in a large range,
e.g., from $a_0=2$ to $a_0=2000$. Such an ion channel may be formed
within a period before it is destroyed by Coulomb explosion, when a
precursor laser pulse or a dense electron beam passes through
plasma.

\begin{acknowledgments}
This work was supported by the National Key R\&D Program of China
(Grant Nos. 2018YFA0404801 and 2018YFA0404802), National Natural
Science Foundation of China (Grant Nos. 11775302, 11721091, and
11520101003), Science Challenge Project of China (Grant Nos.
TZ2016005 and TZ2018005), the Strategic Priority Research Program of
the Chinese Academy of Sciences (Grant No. XDB16010200), and
European Commission H2020-MSCA-IF (Grant No. 743949).
\end{acknowledgments}


\begin{thebibliography}{99}

\bibitem{Danson}C. Danson, D. Hillier, N. Hopps, and D. Neely, High Power Laser Sci. Eng. \textbf{3}, e3 (2015).

\bibitem{SULF_5PW}\url{http://news.sciencenet.cn/htmlnews/2017/10/392187.shtm}

\bibitem{Laser_4PW20fs}\url{https://apri.gist.ac.kr/en/page/menu02/page0101.php}
\bibitem{Laser_1PW_contrast}I. J. Kim, K. H. Pae, I. W. Choi, C.-L. Lee,
H. T. Kim, H. Singhal, J. H. Sung, S. K. Lee, H. W. Lee, P. V.
Nickles, T. M. Jeong, C. M. Kim, and C. H. Nam, Phys. Plasma
\textbf{23}, 070701 (2016).

\bibitem{ELI}\url{http://www.extreme-light-infrastructure.eu}
\bibitem{OMEGA_EP_OPAL}J. D. Zuegel, \emph{Technology Development and Prospects for
100-PW-Class Optical Parametric Chirped-Pulse Amplification Pumped
by OMEGA EP}, plenary talk at the 2nd International Symposium on
High Power Laser Science and Engineering (HPLSE2016), March 15-18,
2016, Suzhou, China. (\url{http://www.hplse.net/dct/page/70005})
\bibitem{XCELS}\url{http://www.xcels.iapras.ru/}



\bibitem{Bell}A. R. Bell and J. G. Kirk, Phys. Rev. Lett. \textbf{101}, 200403 (2008).
\bibitem{Piazza}A. Di Piazza, C. Muller, K. Z. Hatsagortsyan, and C. H.
Keitel, Rev. Mod. Phys. \textbf{84}, 1177 (2012).
\bibitem{Ridgers}C. P. Ridgers, C. S. Brady, R. Duclous, J. G. Kirk, K. Bennett, T.
  D. Arber, A. P. L. Robinson, and A. R. Bell, Phys. Rev. Lett.
  \textbf{108}, 165006 (2012).
\bibitem{Vacuum_QED}Q. Z. Lv, Y. Liu, Y. J. Li, R. Grobe, and Q. Su, Phys. Rev.
Lett. \textbf{111}, 183204 (2013).



\bibitem{Tajima}T. Tajima and J. M. Dawson, Phys. Rev. Lett. \textbf{43}, 267 (1979).
\bibitem{Esarey}E. Esarey, C. B. Schroeder, and W. P. Leemans, Rev. Mod. Phys. \textbf{81},
1229 (2009).
\bibitem{Lu}W. Lu, M. Tzoufras, C. Joshi, F. S. Tsung, W. B. Mori, J. Vieira,
           R. A. Fonseca, L. O. Silva, Phys. Rev. ST Accel. Beams \textbf{10}, 061301 (2007).
\bibitem{Leemans_2014}W. P. Leemans, A. J. Gonsalves, H.-S. Mao, K. Nakamura, C.
Benedetti, C. B. Schroeder, Cs. Toth, J. Daniels, D. E.
Mittelberger, S. S. Bulanov, J.-L. Vay, C. G. R. Geddes, and E.
Esarey, Phys. Rev. Lett. \textbf{113}, 245002 (2014).
\bibitem{Leemans_2019}A. J. Gonsalves, K. Nakamura, J. Daniels, C. Benedetti, C. Pieronek,
T. C. H. de Raadt, S. Steinke, J. H. Bin, S. S. Bulanov, J. van
Tilborg, C. G. R. Geddes, C. B. Schroeder, Cs. Toth, E. Esarey, K.
Swanson, L. Fan-Chiang, G. Bagdasarov, N. Bobrova, V. Gasilov, G.
Korn, P. Sasorov, and W. P. Leemans, Phys. Rev. Lett. \textbf{122},
084801 (2019).


\bibitem{Esirkepov}T. Esirkepov, M. Borghesi, S. V. Bulanov, G. Mourou, and T. Tajima,
Phys. Rev. Lett. \textbf{92}, 175003 (2004).
\bibitem{Yan}X. Q. Yan, C. Lin, Z. M. Sheng, Z. Y. Guo, B. C. Liu, Y.
R. Lu, J. X. Fang, J. E. Chen, Phys. Rev. Lett. \textbf{100}, 135003
(2008).
\bibitem{Bulanov}S. V. Bulanov, E. Yu. Echkina, T. Zh. Esirkepov, I. N. Inovenkov, M.
Kando, F. Pegoraro, and G. Korn, Phys. Rev. Lett. \textbf{104},
135003 (2010).
\bibitem{Macchi}Andrea Macchi, Marco Borghesi, and Matteo Passoni, Rev. Mod.
Phys. \textbf{85}, 751 (2013).
\bibitem{Mima}K. Mima, J. Fuchs, T. Taguchi, J. Alvarez,
J.R. Marques, S.N. Chen, T. Tajima, and J. M. Perlado, Matter and
Radiation at Extremes \textbf{3}, 127 (2018).
\bibitem{Matsui}Ryutaro Matsui, Yuji Fukuda, and Yasuaki Kishimoto, Phys. Rev. Lett.
\textbf{122}, 014804 (2019).

\bibitem{PNAS2018}W.-M. Wang, Z.-M. Sheng, P. Gibbon, L.-M. Chen, Y.-T. Li, J.
Zhang, Proc. Natl. Acad. Sci. USA \textbf{115}, 9911 (2018).
\bibitem{Lei}Bifeng Lei, Jingwei Wang, Vasily Kharin, Matt Zepf, and Sergey
Rykovanov, Phys. Rev. Lett. 120, 134801 (2018).
\bibitem{Chen}Yue-Yue Chen, Jian-Xing Li, Karen Z. Hatsagortsyan, and Christoph H.
Keitel, Phys. Rev. Lett. 121, 074801 (2018).



\bibitem{Esarey97}E. Esarey, P. Sprangle, J. Krall, and A. Ting, IEEE J.
Quantum Electron. \textbf{33}, 1879 (1997).

\bibitem{Channel_exp}A. Butler, D. J. Spence, and S. M. Hooker, Phys. Rev. Lett. \textbf{89},
185003 (2002).


\bibitem{Monot} P. Monot, T. Auguste, P. Gibbon, F. Jakober, G. Mainfray, A. Dulieu,
M. Louis-Jacquet, G. Malka, and J. L. Miquel, Phys. Rev. Lett.
\textbf{74}, 2953 (1995).
\bibitem{Pukhov}A. Pukhov and J. Meyer-ter-Vehn, Phys. Rev. Lett. \textbf{76}, 3975 (1996).


\bibitem{Sun87}G. Z. Sun, E. Ott, Y. C. Lee, and P. Guzdar, Phys. Fluids \textbf{30}, 526 (1987).
\bibitem{Kneip}S. Kneip, S. R. Nagel, C. Bellei, N. Bourgeois, A. E. Dangor, A.
Gopal, R. Heathcote, S. P. D. Mangles, J. R. Marques, A. Maksimchuk,
P. M. Nilson, K. Ta Phuoc, S. Reed, M. Tzoufras, F. S. Tsung, L.
Willingale, W. B. Mori, A. Rousse, K. Krushelnick, and Z. Najmudin,
Phys. Rev. Lett. \textbf{100}, 105006 (2008).
\bibitem{Wang2012}W.-M. Wang, Z.-M. Sheng, M. Zeng, Y. Liu, Z.-D. Hu, S. Kawata, C.-Y. Zheng,
W. B. Mori, L.-M. Chen, Y.-T. Li, and J. Zhang, Appl. Phys. Lett.
\textbf{101}, 184104 (2012).


\bibitem{KLAPS}W.-M. Wang, P. Gibbon, Z.-M. Sheng, and Y.-T. Li, Phys. Rev. E \textbf{91}, 013101 (2015).
\bibitem{Wang-PRE17}W.-M. Wang, P. Gibbon, Z.-M. Sheng, Y.-T. Li, and J. Zhang, Phys. Rev. E \textbf{96}, 013201 (2017).



\bibitem{Zhidkov} A. Zhidkov, J. Koga, A. Sasaki, and M. Uesaka, Phys.
Rev. Lett. \textbf{88}, 185002 (2002).


\bibitem{Whittum}D. H. Whittum, A. M. Sessler, and J. M. Dawson, Phys.
Rev. Lett. \textbf{64}, 2511 (1990).

\bibitem{Werner}P. W. Werner, E. Schamiloglu, J. R. Smith, K. W. Struve, and R. J.
Lipinski, Phys. Rev. Lett. \textbf{73}, 2986 (1994).



\bibitem{ChenXL}X. L. Chen and R. N. Sudan, Phys. Rev. Lett.{\bf 70}, 2082 (1993).
\bibitem{Esarey94}E. Esarey, J. Krall, and P. Sprangle, Phys. Rev. Lett. \textbf{72}, 2887 (1994).
\bibitem{Wang2006}W.-M. Wang and C.-Y. Zheng, Phys. Plasmas {\bf 13}, 053112 (2006).
\bibitem{Sprangle87}P. Sprangle, A. Ting, and C. M. Tang, Phys. Rev. Lett. \textbf{59}, 202 (1987).



\end{thebibliography}
\end{document}